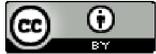

# Models for Quark Stars with Charged Anisotropic Matter

**Manuel Malaver**[1]*

Universidad Marítima del Caribe, Departamento de Ciencias Básicas, Catia la Mar, Venezuela

*Corresponding author (Email: mmf.umc@gmail.com)

**Abstract -** In this paper, we found new exact solutions to the Einstein- Maxwell system of equations within the framework of MIT Bag Model considering a particular form for the measure of anisotropy and a gravitational potential which depends on an adjustable parameter α. Variables as the energy density, radial pressure, tangential pressure, electric field intensity and the metric functions are written in terms of elementary and polinominal functions. We show that the form chosen for the gravitational potential and the anisotropy allows obtain physically acceptable solutions with any value of the adjustable parameter.

## 1. Introduction

One of the fundamental problems in the general theory of relativity is finding exact solutions of the Einstein field equations (Bicak, 2006 and Kuhfitting, 2011). Some solutions found fundamental applications in astrophysics, cosmology and more recently in the developments inspired by string theory (Bicak, 2006). Different mathematical formulations that allow solving Einstein´s field equations have been used to describe the behavior of objects submitted to strong gravitational fields known as neutron stars, quasars and white dwarfs (Malaver, 2013,a; Komathiraj & Maharaj, 2008; Sharma et al., 2001 ).

From the development of Einstein´s theory of general relativity, the description of compact objects has been a central issue in relativistic astrophysics in the last few decades (Bicak, 2006 and Dey et al., 1998). Recent experimental observations in binary pulsars (Dey et al., 1998) suggest that could be quark stars. The existence of quark stars in hydrostatic equilibrium was first by Itoh (1970) in a seminal treatment. Recently, the study of strange stars consisting of quark matter has stimulated much interest since could represent the most energetically favourable state of baryon matter.

In the construction of the first theoretical models of relativistic stars are important the works of Schwarzschild (1916), Tolman (1939) and Oppenheimer & Volkoff (1939). Schwarzschild (1916) found analytical solutions that allowed describing a star with uniform density, Tolman (1939) developed a method to find solutions of static spheres of fluid and Oppenheimer & Volkoff (1939) used Tolman's solutions to study the gravitational balance of neutron stars. It is important to mention Chandrasekhar's contributions (1931) in the model production of white dwarfs in presence of relativistic effects and the works of Baade & Zwicky (1934) that propose the concept of neutron stars and identify an astronomic dense object known as supernovas.

The physics of ultrahigh densities is not well understood and many of the strange stars studies have been performed within the framework of the MIT bag model (Komathiraj & Maharaj, 2007). In this model, the strange matter equation of state has a simple linear form given by $p = \frac{1}{3}(\rho - 4B)$ where $\rho$ is the energy density, p is the isotropic pressure and B is the bag constant. However, theoretical works of realistic stellar models (Herrera & Santos,1997; Cosenza et al., 1981; Gokhroo & Mehra, 1994; Herrera, 1992) it has been suggested that superdense matter may be anisotropic, at least in some density ranges. The existence of anisotropy within a star can be explained by the presence of a solid core, phase transitions, a type III super fluid, a pion condensation (Herrera et al., 1979) or another physical phenomenon. In such systems, the radial pressure is different from the tangential pressure. This generalization has been very used in the study of the balance and collapse of compact spheres (Herrera & de Leon,1985; Herrera & Santos, 1998; Bondi, 1992; Maharaj et al.,2014 ).

Many researchers have used a great variety of mathematical techniques to try to obtain exact solutions for quark stars within the framework of MIT bag model, since it has been demonstrated by Komathiraj & Maharaj (2007), Maharaj et al.(2014), Malaver (2009), Thirukkanesh and Maharaj (2008) and Thirukkanesh and Ragel (2013). Feroze & Siddiqui (2011) and Malaver (2014) consider a quadratic equation of state for the matter distribution and specify particular forms for the gravitational potential and electric field intensity. Mafa Takisa & Maharaj (2013) obtained new exact solutions to the Einstein-Maxwell system of equations with a polytropic equation of state. Thirukkanesh and Ragel (2012) have obtained particular models of anisotropic fluids with polytropic equation of state which are consistent with the reported experimental observations. More recently, Malaver (2013,b,c) generated new exact solutions to the Einstein-Maxwell system considering Van der Waals modified equation of state with and without polytropical exponent. Maharaj et al. (2014) found new models by specifying a particular form for one of the gravitational potentials and the measure of anisotropy. Mak & Harko (2004) found a relativistic model of strange quark star with the suppositions of spherical symmetry and conformal Killing vector.

Using the procedure suggested by Komathiraj & Maharaj (2007) and Maharaj et al. (2014) in the present work, we have generated a new class for charged anisotropic matter with the bag equation of state in static spherically symmetric spacetime and a gravitational potential y(x) of Malaver (2009) that depends on an adjustable parameter α. We have obtained some new classes of static spherically symmetrical models of charged matter considering a particular form for the measure of anisotropy where the variation of the parameter α modifies the radial pressure, tangential pressure, energy density and the electric field intensity of the compact objects. The models of Malaver (2009) and Komathiraj & Maharaj (2007) for isotropic pressure are recovered as special cases. This article is organized as follows, in Section 2, we present Einstein´s field equations. In Section 3 we have particular choices for the gravitational potential y(x) and the anisotropy that allows solving the field equations and we have obtained new models for charged anisotropic matter. In Section 4, a physical analysis of the new solutions is performed. Finally in Section 5, we conclude.

## 2. Einstein Field Equations

Consider a spherically symmetric four dimensional space time whose line element is given in Schwarzschild coordinates by

$$ds^2 = -e^{2\nu(r)} dt^2 + e^{2\lambda(r)} dr^2 + r^2(d\theta^2 + \sin^2\theta d\varphi^2) \quad (1)$$

Using the transformations $x = Cr^2$, $Z(x) = e^{-2\lambda(r)}$ and $A^2 y^2(x) = e^{2\nu(r)}$ with arbitrary constants A and C, suggested by Durgapal & Bannerji (1983), the Einstein field equations as given in (1) are

$$\frac{1-Z}{x} - 2\dot{Z} = \frac{\rho}{C} + \frac{E^2}{2C} \quad (2)$$

$$4Z\frac{\dot{y}}{y} - \frac{1-Z}{x} = \frac{p_r}{C} - \frac{E^2}{2C} \quad (3)$$

$$4xZ\frac{\ddot{y}}{y} + (4Z + 2x\dot{Z})\frac{\dot{y}}{y} + \dot{Z} = \frac{p_t}{C} + \frac{E^2}{2C} \quad (4)$$

$$\sigma^2 = \frac{4CZ}{x}(x\dot{E} + E)^2 \quad (5)$$

Where $\rho$ is the energy density, $p_r$ is the radial pressure, $E$ is electric field intensity, $\sigma$ is the charge density, $p_t$ is the tangential pressure and dot denote differentiation with respect to x.

We can replace the system of fields equations with the bag equation of state by the system

$$\rho = 3p + 4B \quad (6)$$

$$\frac{p}{C} = Z\frac{\dot{y}}{y} - \frac{1}{2}\dot{Z} - \frac{B}{C} \quad (7)$$

$$\frac{E^2}{2C} = \frac{1-Z}{x} - 3Z\frac{\dot{y}}{y} - \frac{1}{2}\dot{Z} - \frac{B}{C} \quad (8)$$

$$p_t = p_r + \Delta \quad (9)$$

$$\Delta = \frac{4xCZ\dot{y}}{y} + C(2x\dot{Z} + 6Z)\frac{\dot{y}}{y} + C\left[2\left(\dot{Z} + \frac{B}{C}\right) + \frac{Z-1}{x}\right] \quad (10)$$

$$\sigma = 2\sqrt{\frac{CZ}{x}}(E + x\dot{E}) \quad (11)$$

The equations (6), (7), (8), (9), (10) and (11) governs the gravitational behavior of a charged quark star. The quantity

$\Delta = p_t - p_r$ is called the measure of anisotropy.

## 3. New Exact Solutions with Anisotropy

Using the method suggested by Komathiraj and Maharaj (2007), it is possible to obtain an exact solution of the Einstein-Maxwell system. In this paper, motivated by Malaver (2009), we take the particular form of the gravitational potential

$$y = (a + \alpha x)^2 \qquad (12)$$

where a is a real constant and $\alpha$ is an adjustable parameter. This potential is regular at the origin and well behaved in the interior of the sphere. Substitution of (12) in (10) allows obtaining the equation of the first order

$$\dot{Z} + \frac{21x^2\alpha^2 + 14x\alpha a + a^2}{8x^2 a\alpha + 6x^3\alpha^2 + 2xa^2} Z = \frac{\left(1 - \frac{2B}{C}x\right)(a^2 + 2\alpha ax + \alpha^2 x^2) + \Delta}{8x^2 a\alpha + 6x^3\alpha^2 + 2xa^2} \qquad (13)$$

We specify the measure of anisotropy of the following form:

$$\Delta = A^2 + \alpha Ax + x^2 \qquad (14)$$

$A$ is a arbitrary constant and $\alpha$ is the adjustable parameter.

We have considered the particular cases for $\alpha = 1, 2$.

For the case $\alpha = 1$, the equation (13) can be written as

$$\dot{Z} + \frac{21x^2 + 14xa + a^2}{8x^2 a + 6x^3 + 2xa^2} Z = \frac{\left(1 - \frac{2B}{C}x\right)(a^2 + 2ax + x^2) + A^2 + Ax + x^2}{8x^2 a + 6x^3 + 2xa^2} \qquad (15)$$

Integrating (15), we obtain

$$Z(x) = \frac{1}{315(x+a)^2(3x+a)} \left[ \begin{array}{l} 3(30x^3 + 21Ax^2 + 84ax^2 + 105a^2x + 35A^2x + 35aAx + 105a^3 + 105aA^2) \\ -\frac{2Bx}{C}(35x^3 + 135ax^2 + 189a^2x + 105a^3) \end{array} \right] \qquad (16)$$

Z(x) allows generate the following analytical model:

$$e^{2\nu} = A^2(a+x)^4 \qquad (17)$$

$$e^{2\lambda} = \frac{315(x+a)^2(3x+a)}{[3(30x^3 + 21Ax^2 + 84ax^2 + 105a^2x + 35A^2x + 35aAx + 105a^3 + 105aA^2)} \\ -\frac{2Bx}{C}(35x^3 + 135ax^2 + 189a^2x + 105a^3)] \qquad (18)$$

$$p_r = \frac{C}{315(x+a)^3(3x+a)} \left[ \begin{array}{l} 6(30x^3 + 21Ax^2 + 84ax^2 + 105a^2x + 35A^2x + 35aAx + 105a^3 + 105aA^2) \\ -\frac{4Bx}{C}(35x^3 + 135ax^2 + 189a^2x + 105a^3) \end{array} \right]$$

$$+ \frac{1}{630(x+a)^2(3x+a)} \left[ \begin{array}{l} 280Bx^3 - 270Cx^2 + 810aBx^2 + 756a^2Bx - 126ACx - 504aCx - 315a^2C \\ -105A^2C + 210a^3B - 105aAC \end{array} \right] \qquad (19)$$

$$- \frac{1}{630(x+a)^3(3x+a)^2} \left[ \begin{array}{l} 6300Bx^5 + 23570aBx^4 + 33732a^2Bx^3 - 567ACx^3 - 2718aCx^3 - 4095a^2Cx^2 \\ -945A^2Cx^2 + 22680a^3Bx^2 - 1260aACx^2 - 4410a^3Cx - 3360aA^2x + 6720a^4Bx \\ +630Ba^5 - 525a^2ACx - 1575a^4C - 1575a^2A^2C \end{array} \right]$$

$$p_t = \frac{C}{315(x+a)^3(3x+a)} \begin{bmatrix} 6(30x^3 + 21Ax^2 + 84ax^2 + 105a^2x + 35A^2x + 35aAx + 105a^3 + 105aA^2) \\ -\frac{4Bx}{C}(35x^3 + 135ax^2 + 189a^2x + 105a^3) \end{bmatrix}$$

$$+ \frac{1}{630(x+a)^2(3x+a)} \begin{bmatrix} 280Bx^3 - 270Cx^2 + 810aBx^2 + 756a^2Bx - 126ACx - 504aCx - 315a^2C \\ -105A^2C + 210a^3B - 105aAC + 18900a^3A^2x^2 + 18900a^3Ax^3 + 5670a^4A^2x \\ +20790aAx^5 + 20790aA^2x^4 + 28980a^2Ax^4 + 630a^5Ax + 28980a^2A^2x^3 + 5670x^7 \\ +5670A^2x^5 + 630a^5A^2 + 5670Ax^6 + 20790ax^6 + 28980a^2x^5 + 18900a^3x^4 + 5670a^4x^3 \\ +630a^5x^2 \end{bmatrix} \quad (20)$$

$$- \frac{1}{630(x+a)^3(3x+a)^2} \begin{bmatrix} 6300Bx^5 + 23570aBx^4 + 33732a^2Bx^3 - 567ACx^3 - 2718aCx^3 - 4095a^2Cx^2 \\ -945A^2Cx^2 + 22680a^3Bx^2 - 1260aACx^2 - 4410a^3Cx - 3360aA^2x + 6720a^4Bx \\ +630Ba^5 - 525a^2ACx - 1575a^4C - 1575a^2A^2C \end{bmatrix}$$

$$\rho = \frac{C}{105(x+a)^3(3x+a)} \begin{bmatrix} 6(30x^3 + 21Ax^2 + 84ax^2 + 105a^2x + 35A^2x + 35aAx + 105a^3 + 105aA^2) \\ -\frac{4Bx}{C}(35x^3 + 135ax^2 + 189a^2x + 105a^3) \end{bmatrix}$$

$$+ \frac{1}{210(x+a)^2(3x+a)} \begin{bmatrix} 2800Bx^3 - 270Cx^2 + 6690aBx^2 + 4956a^2Bx - 126ACx - 504aCx - 315a^2C \\ -105A^2C + 1050a^3B - 105aAC \end{bmatrix} \quad (21)$$

$$- \frac{1}{210(x+a)^3(3x+a)^2} \begin{bmatrix} 6300Bx^5 + 23570aBx^4 + 33732a^2Bx^3 - 567ACx^3 - 2718aCx^3 - 4095a^2Cx^2 \\ -945A^2Cx^2 + 22680a^3Bx^2 - 1260aACx^2 - 4410a^3Cx - 3360aA^2x + 6720a^4Bx \\ +630Ba^5 - 525a^2ACx - 1575a^4C - 1575a^2A^2C \end{bmatrix}$$

$$E^2 = \frac{2C}{315x(x+a)^2(3x+a)} \begin{bmatrix} 855x^3 - 1953ax^2 + 1260a^2x - 63Ax^2 - 105A^2x - 105aAx - 315aA^2 \\ +\frac{2Bx}{C}(35x^3 + 135ax^2 + 189a^2x + 105a^3) \end{bmatrix}$$

$$- \frac{2C}{315(x+a)^3(3x+a)} \begin{bmatrix} 540x^3 + 378Ax^2 + 1512ax^2 + 1890a^2x + 630A^2x + 630aAx + 1890a^3 + 1890aA^2 \\ -\frac{2Bx}{C}(210x^3 + 810ax^2 + 1134a^2x + 630a^3) \end{bmatrix} \quad (22)$$

$$- \frac{1}{315(x+a)^2(3x+a)} \begin{bmatrix} -280Bx^3 + 270Cx^2 - 810aBx^2 - 756a^2Bx + 126ACx + 504aCx + 315a^2C + 105A^2C \\ -210a^3B + 105aAC \end{bmatrix}$$

$$- \frac{1}{315(x+a)^3(3x+a)^2} \begin{bmatrix} 6300Bx^5 + 23570aBx^4 - 810Cx^4 - 2718aCx^3 + 33732a^2Bx^3 + 22680a^3Bx^2 - 567ACx^3 \\ -315aACx^2 - 4095a^2Cx^2 - 4410a^3Cx - 945A^2Cx^2 - 3360aA^2Cx + 6720a^4Bx \\ -945aACx^2 - 525a^2ACx - 1575a^4C - 1575a^2A^2C + 630Ba^5 \end{bmatrix}$$

The metric for this model is

$$ds^2 = -A^2(a+x)^4 dt^2 + \left[ \frac{315(x+a)^2(3x+a)}{4cx\left(3(30x^3 + 21Ax^2 + 84ax^2 + 105a^2x + 35A^2x + 35aAx + 105a^3 + 105aA^2) - \frac{2Bx}{C}(35x^3 + 135ax^2 + 189a^2x + 105a^3)\right)} \right] dx^2 + \frac{x}{c}(d\theta^2 + \sin^2\theta d\varphi^2) \quad (23)$$

With $\alpha=2$, the eq. (13) becomes

$$\dot{Z} + \frac{84x^2 + 28xa + a^2}{16x^2a + 24x^3 + 2xa^2}Z = \frac{\left(1 - \frac{2B}{C}x\right)(a^2 + 4ax + 4x^2) + A^2 + 2Ax + x^2}{16x^2a + 24x^3 + 2xa^2} \quad (24)$$

Integrating (24), we obtain

$$Z(x) = \frac{1}{315(2x+a)^2(6x+a)}\begin{bmatrix} 3(150x^3 + 84Ax^2 + 273ax^2 + 210a^2x + 70A^2x + 70aAx + 105a^3 + 105aA^2) \\ -\frac{2Bx}{C}(280x^3 + 540ax^2 + 378a^2x + 105a^3) \end{bmatrix} \quad (25)$$

With the equation (25), we can generate the exact analytical model

$$e^{2\nu} = A^2(a+2x)^4 \quad (26)$$

$$e^{2\lambda} = \frac{315(2x+a)^2(6x+a)}{\begin{bmatrix} 3(150x^3 + 84Ax^2 + 273ax^2 + 210a^2x + 70A^2x + 70aAx + 105a^3 + 105aA^2) \\ -\frac{2Bx}{C}(280x^3 + 540ax^2 + 378a^2x + 105a^3) \end{bmatrix}} \quad (27)$$

$$p_r = \frac{C}{315(2x+a)^3(6x+a)}\begin{bmatrix} 12(150x^3 + 84Ax^2 + 273ax^2 + 210a^2x + 70A^2x + 70aAx + 105a^3 + 105aA^2) \\ -\frac{4Bx}{C}(560x^3 + 1080ax^2 + 756a^2x + 210a^3) \end{bmatrix}$$

$$+ \frac{1}{630(2x+a)^2(6x+a)}\begin{bmatrix} 2240Bx^3 - 1350Cx^2 + 3240aBx^2 + 1512a^2Bx - 504ACx - 1638aCx - 630a^2C \\ -210A^2C + 210a^3B - 210aAC \end{bmatrix} \quad (28)$$

$$- \frac{1}{630(2x+a)^3(6x+a)^2}\begin{bmatrix} 201600Bx^5 + 377120aBx^4 + 269856a^2Bx^3 - 9072ACx^3 - 33984aCx^3 - 30870a^2Cx^2 \\ -7560A^2Cx^2 + 75600a^3Bx^2 - 10080aACx^2 - 17640a^3Cx - 13440aA^2Cx + 13440a^4Bx \\ +630Ba^5 - 2100a^2ACx - 3150a^4C - 3150a^2A^2C \end{bmatrix}$$

$$p_t = \frac{C}{315(2x+a)^3(6x+a)}\begin{bmatrix} 12(150x^3 + 84Ax^2 + 273ax^2 + 210a^2x + 70A^2x + 70aAx + 105a^3 + 105aA^2) \\ -\frac{4Bx}{C}(560x^3 + 1080ax^2 + 756a^2x + 210a^3) \end{bmatrix}$$

$$+ \frac{1}{630(2x+a)^2(6x+a)}\begin{bmatrix} 2240Bx^3 - 1350Cx^2 + 3240aBx^2 + 1512a^2Bx - 504ACx - 1638aCx - 630a^2C \\ -210A^2C + 210a^3B - 210aAC + 15120A^2x^3 + 17640aA^2x^2 + 6300a^2A^2x + 630a^3A^2 \\ +30240x^4 + 35280aAx^3 + 12600a^2Ax^2 + 1260a^3Ax + 15120x^5 + 17640ax^4 + 6300a^2x^3 \\ +630a^3x^2 \end{bmatrix} \quad (29)$$

$$- \frac{1}{630(2x+a)^3(6x+a)^2}\begin{bmatrix} 201600Bx^5 + 377120aBx^4 + 269856a^2Bx^3 - 9072ACx^3 - 33984aCx^3 - 30870a^2Cx^2 \\ -7560A^2Cx^2 + 75600a^3Bx^2 - 10080aACx^2 - 17640a^3Cx - 13440aA^2Cx + 13440a^4Bx \\ +630Ba^5 - 2100a^2ACx - 3150a^4C - 3150a^2A^2C \end{bmatrix}$$

$$\rho = \frac{3C}{315(2x+a)^3(6x+a)} \begin{bmatrix} 12(150x^3+84Ax^2+273ax^2+210a^2x+70A^2x+70aAx+105a^3+105aA^2) \\ -\frac{4Bx}{C}(560x^3+1080ax^2+756a^2x+210a^3) \end{bmatrix}$$

$$+\frac{1}{630(2x+a)^2(6x+a)}\begin{bmatrix} 67200Bx^3-4050Cx^2+80280aBx^2+29736a^2Bx-1512ACx-4914aCx-1890a^2C \\ -630A^2C+3150a^3B-630aAC \end{bmatrix} \quad (30)$$

$$-\frac{1}{210(2x+a)^3(6x+a)^2}\begin{bmatrix} 201600Bx^5+377120aBx^4+269856a^2Bx^3-9072ACx^3-33984aCx^3-30870a^2Cx^2 \\ -7560A^2Cx^2+75600a^3Bx^2-10080aACx^2-17640a^3Cx-13440aA^2Cx+13440a^4Bx \\ +630Ba^5-2100a^2ACx-3150a^4C-3150a^2A^2C \end{bmatrix}$$

$$E^2 = \frac{2C}{315x(2x+a)^2(6x+a)}\begin{bmatrix} 7110x^3+8001ax^2+2520a^2x-252Ax^2-210A^2x-210aAx-315aA^2 \\ +\frac{2Bx}{C}(280x^3+540ax^2+378a^2x+105a^3) \end{bmatrix}$$

$$-\frac{2C}{315(2x+a)^3(6x+a)}\begin{bmatrix} 5400x^3+3024Ax^2+9828ax^2+7560a^2x+2520A^2x+2520aAx+3780a^3+3780aA^2 \\ -\frac{2Bx}{C}(3360x^3+6480ax^2+4536a^2x+1260a^3) \end{bmatrix} \quad (31)$$

$$-\frac{1}{315(2x+a)^2(6x+a)}\begin{bmatrix} -2240Bx^3+1350Cx^2-3240aBx^2-1512a^2Bx+504ACx+1638aCx+630a^2C+210A^2C \\ -210a^3B+210aAC \end{bmatrix}$$

$$-\frac{1}{315(2x+a)^3(6x+a)^2}\begin{bmatrix} 201600Bx^5+377120aBx^4-33984aCx^3+269856a^2Bx^3+75600a^3Bx^2-9072ACx^3 \\ -10080aACx^2-30870a^2Cx^2-17640a^3Cx-7560A^2Cx^2-13440aA^2Cx+13440a^4Bx \\ -2100a^2ACx-3150a^4C-3150a^2A^2C+630Ba^5 \end{bmatrix}$$

For this model, the metric is

$$ds^2 = -A^2(a+2x)^4 dt^2 + \left[\frac{315(2x+a)^2(6x+a)}{4cx\begin{pmatrix} 3(150x^3+84Ax^2+273ax^2+210a^2x+70A^2x+70aAx \\ +105a^3+105aA^2)-\frac{2Bx}{C}(280x^3+540ax^2+378a^2x+105a^3) \end{pmatrix}}\right] dx^2 + \frac{x}{c}(d\theta^2+\sin^2\theta d\phi^2) \quad (32)$$

## 4. Physical Analysis

The presented models constitute another new family of solutions for a charged quark star with anisotropic pressure. The metric functions can be written in terms of polinominal functions, and the variables energy density, pressure and charge density also are represented analytical. For the case $\alpha=1$, the metric functions $e^{2\nu(r)}$ and $e^{2\lambda(r)}$ behaves well inside the star and have a finite value of $e^{2\nu(r)}=a^4 A^2$ and $e^{2\lambda(r)}=\frac{a^2}{a^2+A^2}$ at the center x=0. The energy density is positive in the interior and in the center takes the value $\rho=12\frac{C}{a}+2B+\frac{40CA^2-aAC}{2a^3}$. The pressure $p_r$ is regular and in the center is $p_r=4\frac{C}{a}-\frac{2B}{3}+\frac{26CA^2-aAC}{6a^3}$. The tangential pressure $p_t$ also is regular and continuous and in the center x=0 take the value $p_t=4\frac{C}{a}-\frac{2B}{3}+\frac{26CA^2+6a^5A^2-aAC}{6a^3}$. This class of solutions has a singularity in the electrical field and the charge density. When $\Delta=0$, we obtain the Komathiraj and Maharaj model (2007) for isotropic pressure as a special case. For the solution found with $\alpha=2$, the functions $e^{2\nu(r)}$ and $e^{2\lambda(r)}$ acquire finite values in the center x = 0 as in the case for $\alpha = 1$. The energy density, the radial pressure and the tangential pressure take the values $\rho = 24\dfrac{C}{a} + 2B + \dfrac{26CA^2 - aAC}{a^3}$, $p_r = 8\dfrac{C}{a} - \dfrac{2B}{3} + \dfrac{26CA^2 - aAC}{3a^3}$, $p_t = 8\dfrac{C}{a} - \dfrac{2B}{3} + \dfrac{26CA^2 + 3a^3A^2 - aAC}{3a^3}$. Again, as in the case $\alpha = 1$, there is a singularity in the electrical field and the charge density. For isotropic pressure $\Delta = 0$ and we recover Malaver´s model (2009) with $\alpha = 2$ as an especial case.

## 5. Conclusion

We have generated a new class of exact solutions for the Einstein-Maxwell system within the framework of MIT Bag Model with a particular form for the measure of anisotropy. We have studied two new types of analytical solutions specifying the form of the gravitational potential $y(x)$ which depends on an adjustable parameter α. Both solutions correspond to models which have finite values for the energy density, the radial pressure and the tangential pressure at the center of the star but present singularities in the charge density and electric field. This singularity in x=0 not appear in the energy density that always remains finite, which contrasts with the Mak and Harko model (2004). The proposed model allows us to recover the solutions with isotropic pressures as a special case. The method to generate analytical models with anisotropy will depend on the form of and $y(x)$ and $\Delta$, necessary to determine physically acceptable solutions.